\documentclass[useAMS,usenatbib]{mn2e}
\usepackage{amsmath}
\usepackage[toc,page]{appendix}
\usepackage{pslatex}
\usepackage{amsmath}

\usepackage{graphicx}
\usepackage{float}

\title[Thermal surface waves in a protoplanetary disk]{Simulations of thermal surface
waves in a protoplanetary disk using 1+1D approximation}
\author[Pavlyuchenkov et. al.]{Ya. N. Pavlyuchenkov, L. A. Maksimova, V. V. Akimkin \\
        Institute of Astronomy RAS, Moscow, Russia}
\date{}
\begin{document}
\date{Received 30.10.2021; revised 17.01.2022; accepted 24.01.2022 \\
pavyar@inasan.ru}


\maketitle

\begin{abstract}
\noindent
Heating by the central star is one of the key factors determining the
physical structure of protoplanetary disks. Due to the large optical
thickness in the radial direction, disk midplane regions are heated by
the infrared radiation from the disk surface (atmosphere), which in turn
is directly heated by the star. It was previously shown that interception
of the stellar radiation by inhomogeneities on the disk surface can cause
perturbations that propagate towards the star. In this work, we
investigate the occurrence of such waves within a detailed 1+1D numerical
model of the protoplanetary disk. We confirm the previous findings that
in the disk, that is optically thick to its own radiation, the surface
perturbations indeed occur and propagate towards the star. However,
contrary to some analytical predictions, the thermal waves in
sufficiently massive disks affect only the upper layers without
significant fluctuations of temperature in the midplane. Our results
indicate the need to study this instability within more consistent
hydrodynamic models. \newline\newline {\bf DOI:}
10.31857/S000462992205005X
\end{abstract}

\section{Introduction}

Solving the problem of the formation of stars and planets is an important
astrophysical and worldview challenge. However, despite the success in
building a general pictures of stars and planets formation, details of
the protoplanetary disks evoltion (precursors of planetary systems) and
the role of various physical processes have not yet been fully clarified.
This is primarily due to the variety of physical processes in
protoplanetary disks and their complex interaction. In protoplanetary
disks, the conditions for occurrence of a wide variety of dynamic
instabilities are satisfied. Their development can affect both the
observational manifestations and the general disk evolution.

Recent observations made with the ALMA telescope array have shown that
the disk surface density distributions do not correspond to smooth power
laws often used by theorists. On the contrary, on the scale of tens to
hundreds of au bright rings and dark gaps are common phenomena
(\cite{2015ApJ...808L...3A, 2020ARA&A..58..483A, 2018ApJ...869L..42H}).
For now, these rings and gaps are most often explained by the influence
of invisible planets (\cite{2014prpl.conf..667B,
2015ApJ...809...93D,2016MNRAS.459L...1D,
2017ApJ...850..201B,2018ApJ...869L..47Z}). However, many alternative
scenarios have been proposed to explain these features. One of them is
the instability associated with the interception of stellar radiation by
the surface inhomogeneities of the disk.

This instability (known as irradiation instability or thermal wave
instability) can cause the emergence of surface waves traveling towards
the star, and ring structures in circumstellar disks
(\cite{1999ApJ...511..896D, 2000A&A...361L..17D,2008ApJ...672.1183W,
2012A&A...539A..20S}). The mechanism of this physical instability is as
follows. If a small bump forms on the surface of the disk, then the
illuminated side of the bump, facing the star, receives more starlight
and heats up. The warm elements of the bump heat the lower layers of the
disk with their radiation, as a result the formed bump grows. Since the
side of the bump facing the star heats up more strongly than the opposite
side, the perturbation also begins to move towards the star.

Recently, significant progress has been made in the study of this
instability, see \cite{2021ApJ...914L..38U} and
\cite{2021ApJ...923..123W}. In particular, in \cite{2021ApJ...923..123W}
within the framework of an analytical approximation, it is shown that
this instability actually takes place. Moreover, the key role in the
development of instability is the detailed account of dependence of $H/h$
on the distance (where $H$ is the optical height of the disk, and $h$ is
the characteristic hydrostatic scale height). In addition to the
analytical model, the authors of~\cite{2021ApJ...923..123W} also
presented semi-analytical model where the evolution of the midplane
temperature is derived using the heating function calculated by the
radiation transfer code RADMC-3D.

The purpose of this work is to study the irradiation instability
within a more detailed numerical model of the protoplanetary disk,
where the transfer of stellar radiation is treated in two-dimensional
approach while the non-stationary thermal disk structure
is calculated on the entire vertical scale of the disk.

\section{Quasi-hydrostatic 1+1D model of a protoplanetary disk}

\begin{figure*}
\includegraphics[width = 0.49\linewidth]{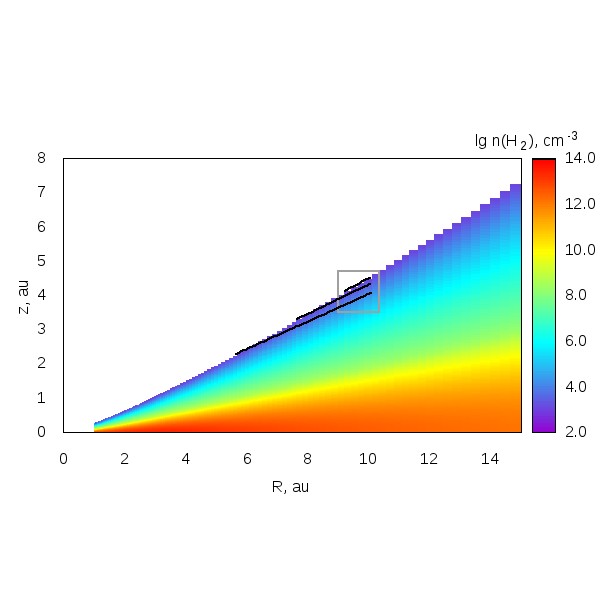}
\includegraphics[width = 0.49\linewidth]{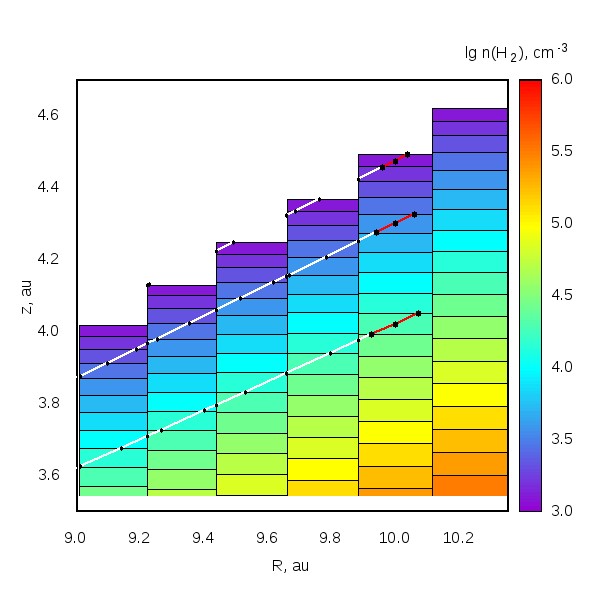}
\caption{Illustration of the ray tracing on the adopted discrete
grid. When integrating the radiation transfer equation along the ray
from the star to the current cell, all intersections of the ray
with the cell boundaries are found (shown as black dots in the right
panel), which is used for accurate calculation of total optical depth.
The final segments of rays belonging to the cells in which heating
is calculated are highlighted in red. }
\label{fig_0a}
\end{figure*}

\begin{figure*}
\includegraphics[width=1\columnwidth]{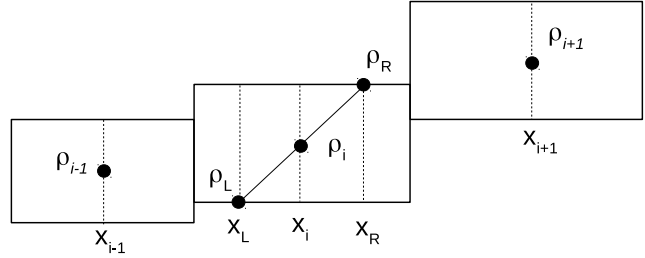}
\caption{Diagram explaining the calculation of the heating function
in the cell.}
\label{fig_0b}
\end{figure*}

To calculate the evolution of the disk, we adopt the model described in
\cite{2017A&A...606A...5V, 2020ARep...64....1P, 2020ARep...64..815M}.
This model solves radiative transfer problem, taking into account the
heating by stellar and interstellar radiation, as well as the diffusion
of thermal (infrared) radiation of the disk itself. The vertical disk
structure is derived consistently with the temperature calculation.
Diffusion of thermal radiation in the model is calculated only in the
vertical ($z$) direction. To model it, we solve system of moment
transport equations in the Eddington approximation:
\begin{eqnarray}
\label{m1}
&&c_{\rm V} \dfrac{\partial T}{\partial t} 
= \kappa_{\rm P} c (E-aT^4) + S \label{m1} \\
&&\dfrac{\partial E}{\partial t} - \dfrac{\partial}{\partial z}
\left(\dfrac{c}{3\rho\kappa_{\rm R}} \dfrac{\partial E}{\partial z}\right) 
= -\kappa_{\rm P} c(E-aT^4),
\label{m2}
\end{eqnarray}
where $T$ is the medium temperature, $E$ is the density of radiation
energy, $z$ is the vertical coordinate, $\rho$ is the gas+dust volume
density, $c_{V}$ is the heat capacity of the gas-dust medium, $c$ is the
speed of light, $a$ is the radiative constant, $\kappa_{\rm P}$ [cm$^2$
g$^{-1}$] is the Planck-averaged absorption coefficient (per unit mass of
the gas-dust mixture), $\kappa_{\rm R}$ [cm$^2$ g$^{-1}$] is the
Rosseland-averaged extinction coefficient (absorption+scattering), and
$S$ [erg s$^{-1}$ g$^{-1}$] is the heating function (per unit mass of
gas-dust medium) by the stellar and interstellar radiation, $S=S^*+S_{\rm
bg}$.

The UV intensity required for the calculation of the heating function
$S^*$ is found for each cell by the direct integration of the radiative
transfer equation from the star to the considered element throughout the
entire disk. This two-dimensional procedure differs from the method
described in \cite{2017A&A...606A...5V}, where the angle between the
direction to the star and the surface of the disk was assumed to be
constant, and thus the problem there was reduced to one-dimensional one.
In the modified method, when integrating transport equation along the ray
all intersections with the cell boundaries are identified (see
Fig.~\ref{fig_0a}), which is used for exact calculation of the optical
depth.

The key aspect in calculating the heating function by stellar
radiation within our model is the consideration of the
radial density gradient inside the cell. In our model, the heating
function $s^{*}$ [erg cm$^{-3}$ s$^{-1}$] by the stellar radiation is
calculated as follows:
\begin{equation}
s^{*}=\rho^{*} \kappa
\frac{L\exp{(-\tau)}}{4\pi R^2}\, \left(\frac{1-\exp{(-\Delta\tau})}{\Delta\tau} \right)
\label{interpol}
\end{equation}
where $L$ is the luminosity of the star, $\kappa$ [cm$^2$ g$^{-1}$] is the
absorption coefficient for stellar radiation, $R$ is the radial distance from
the star to the cell center, $\tau$ is the the total optical depth along the ray
up to the point of entry into the cell, $\Delta \tau = \kappa
\rho^{*} \Delta l$ is the optical thickness of the cell itself along the
ray, $\Delta l$ is the length of the ray segment inside the cell,
$\rho^{*}=\dfrac{1}{4}\left(\rho_{L}+2\rho_{i}+\rho_{r}\right)$ is the
average density along the ray. For the deriving of the
Equation~\eqref{interpol} from the formal solution of the radiative transfer
equation we assumed that the density along the ray inside the cell
changes linearly from $\rho_{L}$ to $\rho_{i}$ and from $\rho_{i}$ to
$\rho_{R}$, see diagram in Fig.~\ref{fig_0b}. The values $\rho_{L}$ and
$\rho_{R}$, in turn, are found using linear density interpolation
between the center of the current cell and the centers of adjacent left and
right cells:
\begin{eqnarray}
&&\rho_{L}=\rho_{i}+\frac{x_{i}-x_{L}}{x_{i}-x_{i-1}}\left(\rho_{i-1}-\rho_{i}\right)\\
&&\rho_{R}=\rho_{i}+\frac{x_{R}-x_{i}}{x_{i+1}-x_{i}}\left(\rho_{i+1}-\rho_{i}\right).
\end{eqnarray}

Thus, when calculating the heating function, we use the density averaged
over neighboring radial cells rather than the central value (which is
adopted in the procedure for calculating the vertical disk structure and
assumed to be constant inside the cell). In the absence of such (or more
precise) averaging, the cell with larger $R$ does not affect cells with
smaller $R$, and as a consequence, there will be no mechanism for the
emergence of thermal wave traveling from outside to inside. With
accounting for the density interpolation, the potential mechanism for the
appearance of irradiation instability is the following. Let us consider
two adjacent columns, inner and outer. Let the outer column heats up, so
the density $\rho_{i+1}$ in its upper layers increases, because the
characteristic scale of the disk height increases. Increased density in
the outer cells leads to an increase in the average density $\rho^{*}$ in
adjacent inner cells, which leads to an increase of $s^{*}$. Further
development of the process depends on the characteristic thermal time
scale and geometric effects of the radiation interception by the inner
column.

Heating by the interstellar UV radiation is calculated as:
\begin{equation}
S_{\rm bg} = \kappa_{\rm uv} W\,\sigma T^4 \exp{(-\tau)} \left(\frac{1-\exp{(-\Delta\tau})}{\Delta\tau} \right),
\end{equation}
where $\kappa_{\rm uv}$ is the absorption coefficient for the interstellar
radiation, $W=10^{-14}$, $T_{\rm bg}=10^4$~K are the dilution and temperature
of the interstellar radiation, $\tau$ is the optical depth from the surface disk to
the current volume element along the vertical direction, $\Delta\tau$ is the
optical thickness of the current element, and $\sigma$ is the
Stefan-Boltzmann constant. Heating by the interstellar radiation is included in
the model, since it can be comparable to the heating from the star in the
outer parts of the disk at the moments of eclipses of the star by the
emerging inhomogeneities of disk surface.

The solution of equations \eqref{m1}--\eqref{m2} is found using a fully
implicit scheme, which is similar to the numerical scheme for solving the
quasilinear equation of thermal conductivity with variable coefficients,
described in the book \cite{Kalitkin}. With linearizing equations and
using Newton's method, iterations can diverge at a sufficiently large
time step. This situation is rare but still occurs. When divergence of
the iterative Newton process appears, we split the time step into
substeps. The method allows us to calculate thermal evolution in all disk
locations, including optically thick regions in which the characteristic
times of heating and cooling processes are comparable with dynamic times.

The thermal structure calculation in this model is closely
related to the restoration of the vertical disk structure
under the assumption of the local hydrostatic equilibrium, which is found
from the following equation:
\begin{equation}
\frac{k_{B}}{\mu_{\rm g} m_{\rm H}}
\dfrac{d(\rho T)}{\rho\, dz} =  -\dfrac{G M_\ast}{r^3}z,
\label{static}
\end{equation}
where $k_{B}$ is the Boltzmann constant, $\mu_{\rm g}=2.3$ is the mean
molecular weight, $m_{\rm H}$ is the atomic mass unit, $G$ is the gravitational
constant, and $M_\ast$ is the mass of the star. Note that when restoring the
vertical structure, we do not take into account the self-gravity of
the disk, which is justified for the disk parameters used in this work. The
calculation of the vertical structure makes it possible to
obtain complete information about the distribution of the density and
temperature in the disk. For solving the equation of hydrostatic
equilibrium an implicit method is also used.

The important condition for the effectiveness of these methods is the
optimal choice of a spatial grid in the $z$-direction. The spatial grid
should resolve all a priori unknown features of the solution
(density and temperature gradients) taking into account significant
limitation on the number of cells and large variation of gas density
(up to 10 orders of magnitude). We have developed an algorithm
for the construction and adaptive modification of such a grid, based on
approximate fast solution of the equation of hydrostatic equilibrium.
This method of restoring the disk vertical structure  with the
calculation of radiation transfer has been extensively tested and compared
with other methods. In the stationary mode, the temperature distribution
is in good consistence with the results of disk structure modeling
obtained by other authors. In the non-stationary regime, the
characteristic times of arrival to thermal equilibrium correspond to
analytical estimates. More details with description of this method can be
found in the article by~\cite{2017A&A...606A...5V}.

The model assumes that the only source of opacity is dust, and the
temperatures of gas and dust are equal. Dust to gas mass ratio
is assumed to be constant throughout the disk and equal to 0.01. A
feature of the thermal model is the use of Planck and Rosseland
temperature-dependent opacities. These coefficients are taken from
~\cite{2020ARep...64....1P}, where they are
described in detail.

\begin{figure*}
\includegraphics[width = \linewidth]{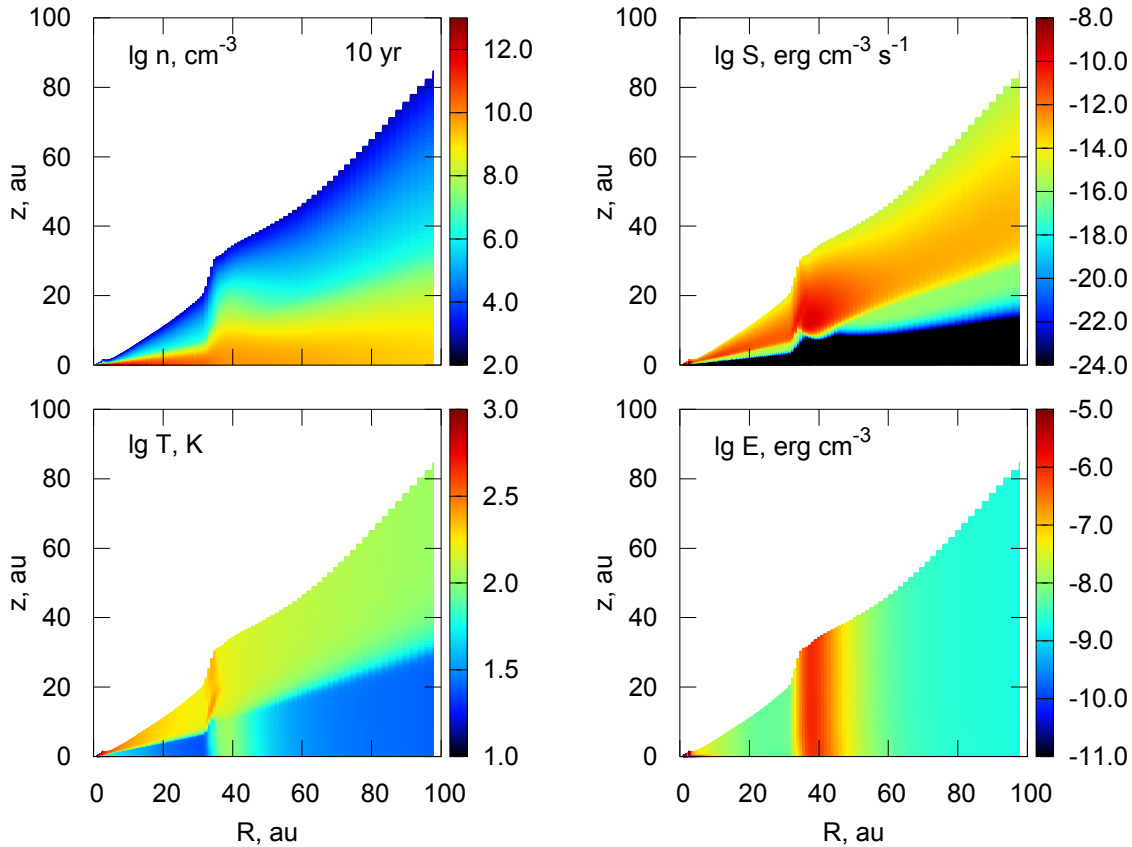}
\caption{The structure of the protoplanetary disk with
$\Sigma_0=10^2$~g~cm$^{-2}$ at 10~yr after the start of evolution. Shown
are the distribution of the logarithm of the gas number density (top
left), the logarithm of the temperature (bottom left), the logarithm of
the UV heating function (top right), and the logarithm of the IR energy
density.}
\label{fig_01}
\end{figure*}

The main input parameters of the model are the mass and luminosity of the
star, which we assume to be solar, as well as the distribution of the
surface density, which is chosen by us in the form:
\begin{equation}
\Sigma(R) = \Sigma_0 \left(1-e^{-\left(\frac{R}{R_0}\right)^{p}} \right)
 \left(\frac{R}{R_{au}}\right)^{q},
\label{dens}
\end{equation}
where $\Sigma_0$ is the surface density near the inner boundary
$R_{au}$=1~au, $q=-1$ is the slope of the density distribution,
$R_{0}$=3~au, $p=8$ is the density distribution smoothing parameters near
the inner edge of the disk. The inner and outer boundaries of the disk
are equal to 1 and 100~au, respectively. Note that in the absence of
smoothing the surface density distribution near the inner disk boundary,
the inner cells intercept most of the stellar radiation and strongly
influence the structure and evolution of the disk, making it difficult to
analyze the surface instability itself. Initial disk state is calculated
under the assumption of a constant angle of entry of radiation into the
disk within the thermal model \cite{2017A&A...606A...5V}. The time step
is chosen constant and equal to $0.1$~yr, which is less than the
characteristic thermal time scales at the adopted model parameters. The
calculation is carried out on a grid of 200 radial $\times$ 120 vertical
cells.

The difference between the presented numerical model and analytical
model used in \cite{2021ApJ...923..123W} is
the treatment of the detailed two-dimensional structure of the disk.
In our case, the heating by the stellar UV radiation is calculated using
direct integration of radiation transfer equation, while the intristic
diffusive thermal radiation is simulated over whole vertical extent.

\section{Results}

Modeling the disk evolution within the framework of the described 1+1D
approach shows that perturbations spontaneously arise in the disk,
propagating towards the star. The characteristics of these perturbations
strongly depend, in particular, on the assumed surface density of the
disk. Consider simulation results for a disk with
$\Sigma_0=10^2$~g~cm$^{-2}$ (corresponding to the disk mass $7\times
10^{-3}$~M$_\odot$). Fig.~\ref{fig_01} shows the disk structure
(distributions of the gas number density, temperature, UV heating and
energy density of IR radiation) at the moment 10~yr after the start disk
evolution. In the vicinity of 35~au one can see a bend in the surface
geometry, behind which the characteristic scale height of the disk
increases. This perturbation propagates inward. Behind the bend, the
stellar radiation heating function increases in the disk atmosphere,
which is associated with a steeper angle of stellar radiation
entry into the disk. The maximum of the heating function in the bend
vicinity is associated with the temperature maximum. The temperature is
also increased in the disk midplane at some distance behind the bend,
which is due to finite time scale for heating of the inner layers by IR
radiation of the surface layers. In the vicinity of the perturbed layer
the energy density of infrared radiation is enhanced, which also
indicates stronger heating of this layer.

\begin{figure*}
\includegraphics[width = 0.49\linewidth]{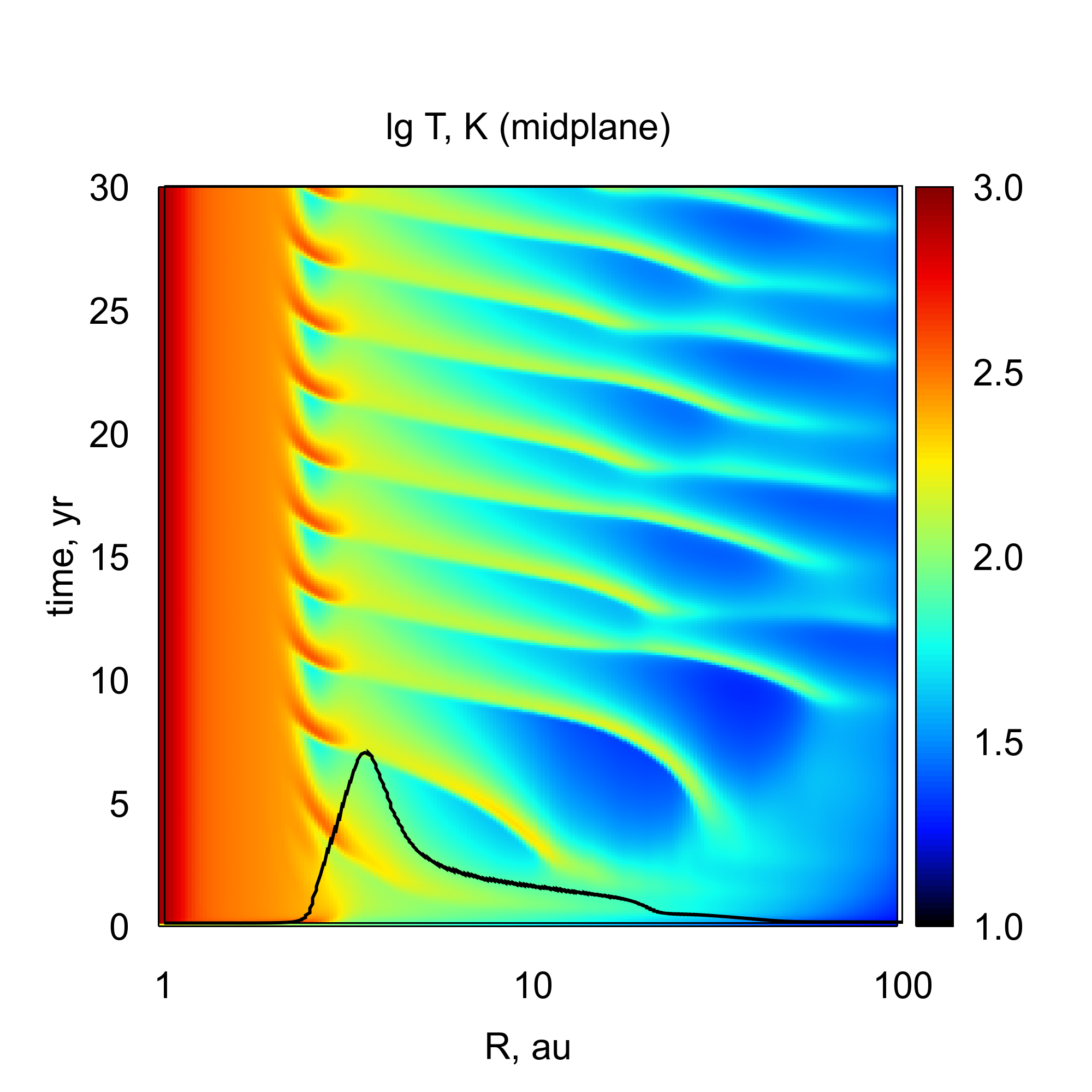}
\includegraphics[width = 0.49\linewidth]{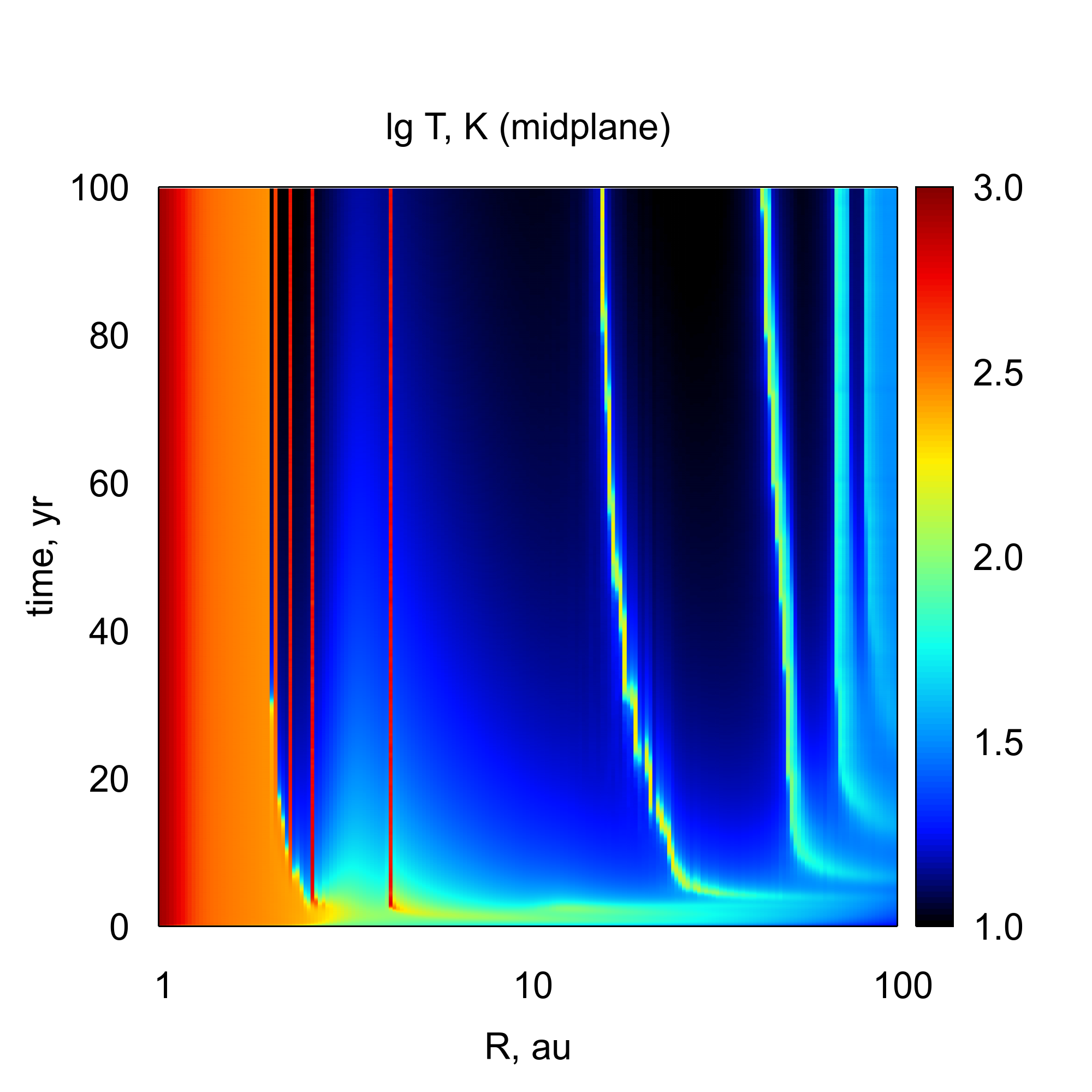}
\caption{
Left panel: evolution of midplane temperature $T_{\rm m}$ 
for the disk model with $\Sigma_0=10^2$~g~cm$^{-2}$.
The distance to the star is shown on the horizontal axis,
time is along the vertical axis.
The black line shows the dependence of the characteristic
thermal time scale $t_{\rm therm}$ on the distance for a time moment of
30~yr. Right panel: evolution of $T_{\rm m}$ for a model without
interpolation of density in the radial direction in the calculation of the
heating rate by the stellar radiation.}
\label{fig_02}
\end{figure*}

The space-time diagram showing the evolution of the midplane temperature
is shown in the left panel of Fig.~\ref{fig_02}.
Inward moving waves stand out well on it, they correspond to maxima
moving from right to left with increasing time. Waves extend from the
outer edge of the disk to the region of $\approx3$~au, which is
transparent to the UV radiation of the star (due to its low density because
of the smoothing of the $\Sigma(R)$ distribution). This diagram also includes
the distribution of the characteristic thermal time calculated as:
\begin{equation}
t_{\rm therm} =\frac{3}{8}\frac{c_{p}\,\Sigma\,\tau_{\rm IR}}{\sigma T_{\rm m}^3},
\label{therm}
\end{equation}
where $\tau_{\rm IR} = \kappa_{P}\Sigma$ is the optical depth in vertical
direction for intrinsic thermal radiation, $T_{\rm m}$ is the midplane
temperature, $c_p$ is the specific heat, and $\sigma$ is the Stefan--Boltzmann
constant. Equation~\eqref{therm} is obtained under assumption that the disk is
optically thick to its own thermal radiation, and serves as an estimate
for the characteristic heating (or cooling) time scale of the midplane layers
to the temperature of $T_{\rm m}$. Periodicity of waves of $\approx3$~yr is
close to the maximum characteristic thermal time scale $\approx7$~yr, i.e.
the perturbed regions of the disk do warm up to the midplane.

The right panel of Fig.~\ref{fig_02} shows the evolution of the
midplane temperature for a disk model with the same parameters, but
without interpolation of the density over the cell in radial direction
during the calculation of the heating rate by the stellar radiation, i.e. where
$\rho^{*}=\rho_{i}$ is assumed, see Equation~\eqref{interpol}. In this case,
several bumps appear in the disk, intercepting the radiation of the
star and shading areas behind them. The slow movement of these humps
towards the star at the initial stages of the disk evolution is associated with the
establishing the self-consistent solution over the thermal time scale.
Regular surface waves do not arise in such a model. This simulation shows
the importance of using high approximation schemes
when integrating the radiative transfer equation in this problem.

\begin{figure*}
\includegraphics[width = \linewidth]{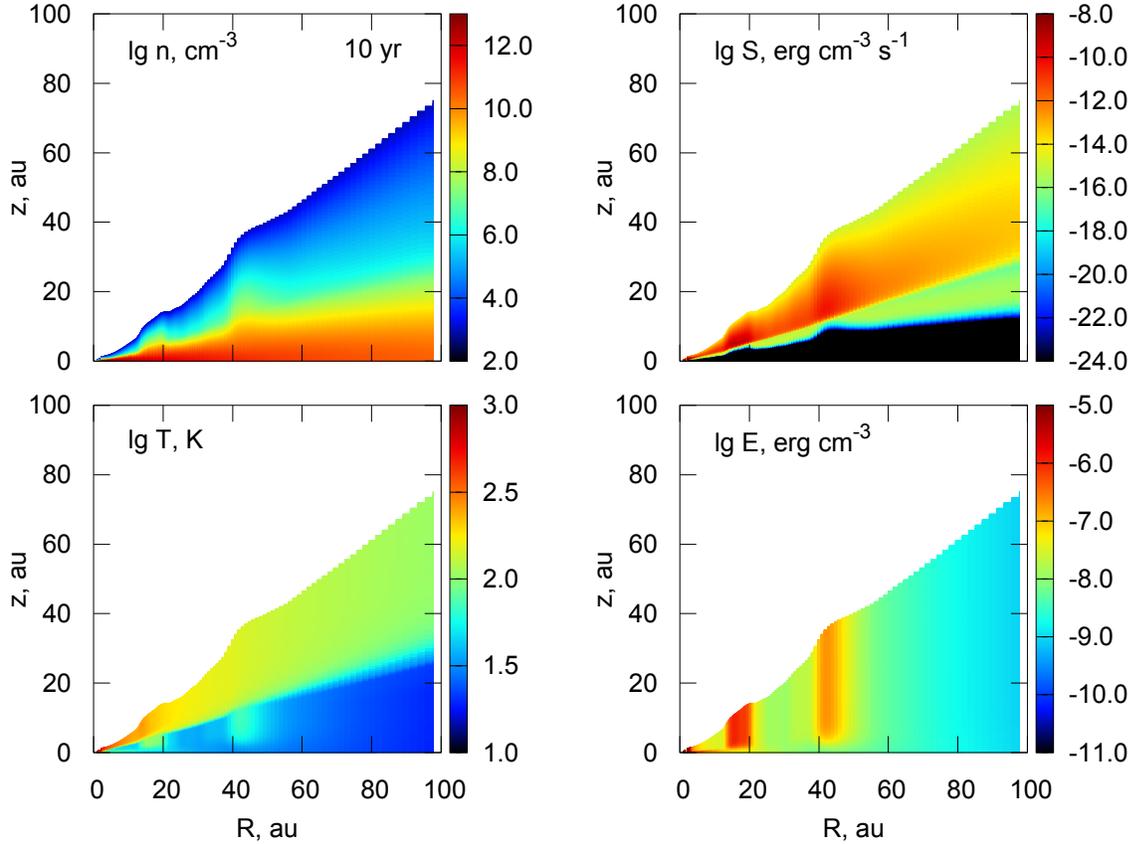}
\caption{
Disk model for $\Sigma_0=10^3$~g~cm$^{-2}$ at 10 yr after the start of evolution.}
\label{fig_03}
\end{figure*}

\begin{figure*}
\includegraphics[width = 0.49\linewidth]{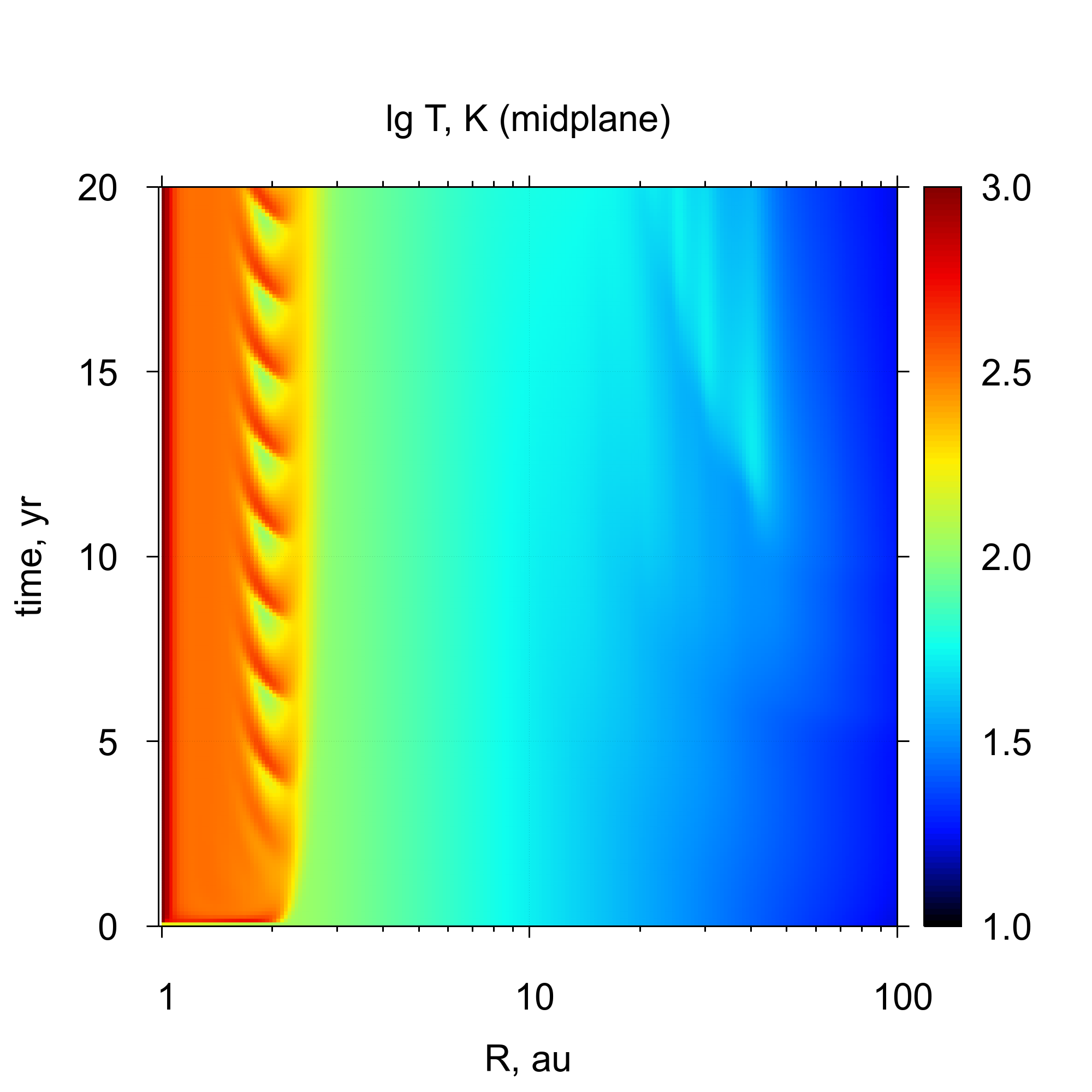}
\includegraphics[width = 0.49\linewidth]{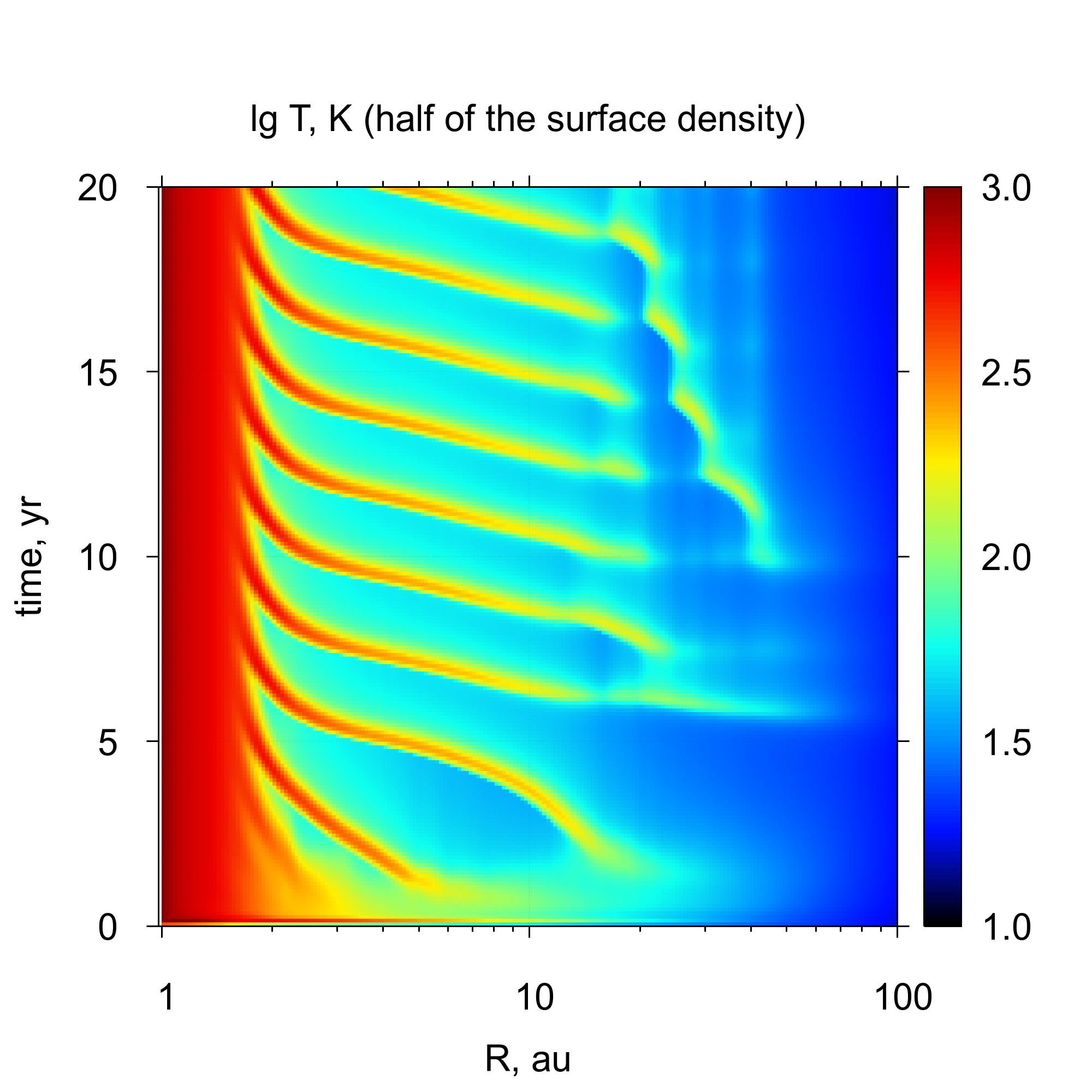}\\
\includegraphics[width = 0.49\linewidth]{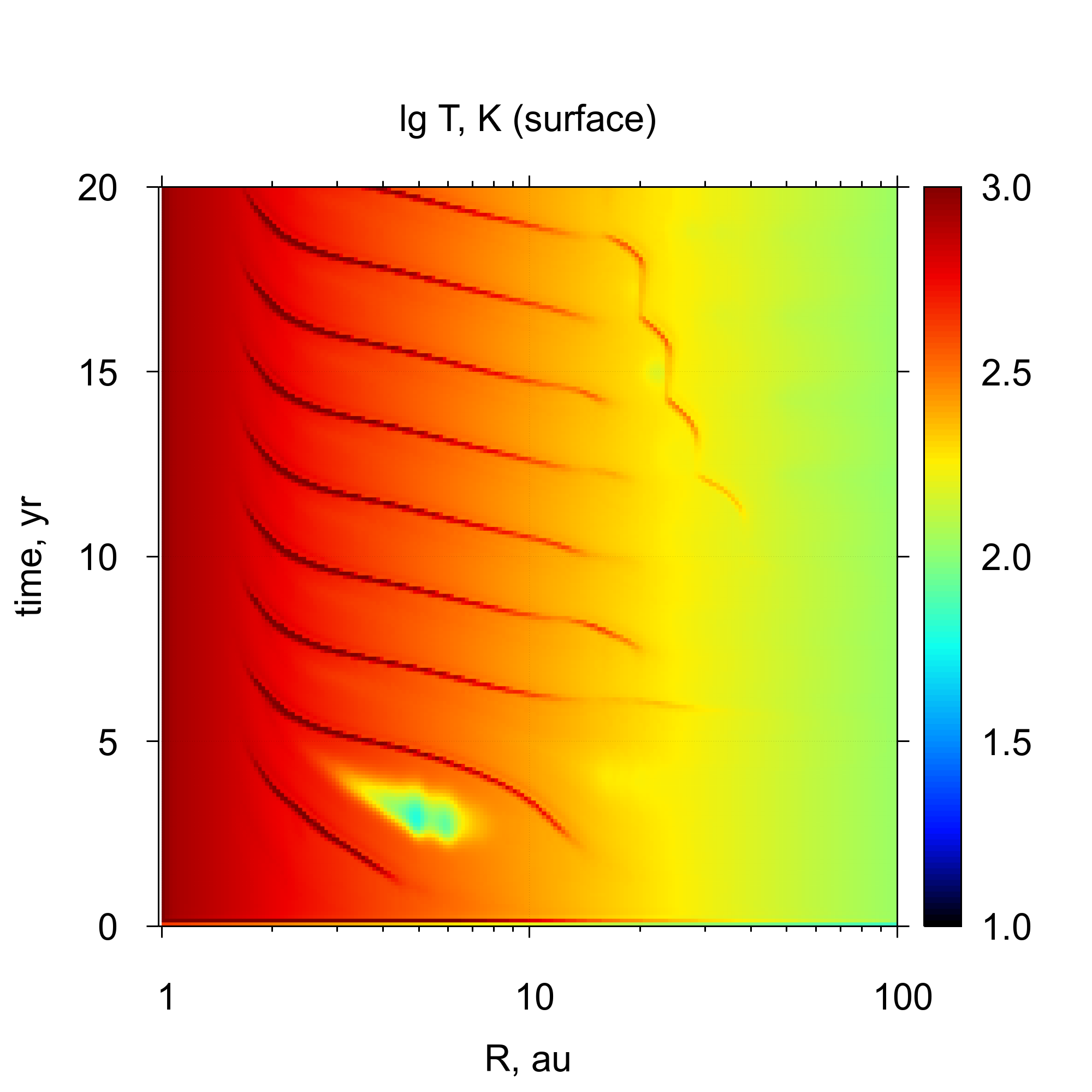}
\includegraphics[width = 0.49\linewidth]{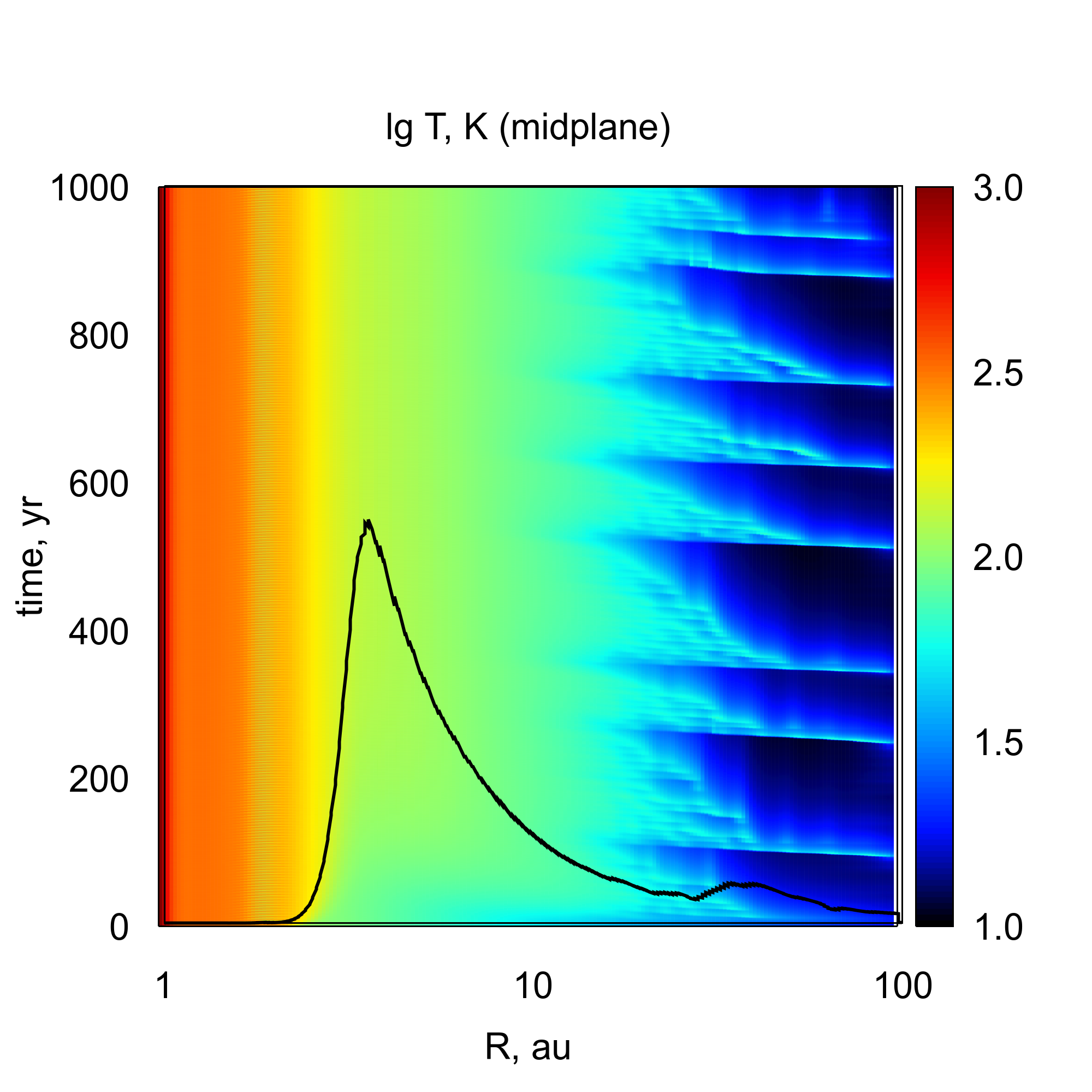}
\caption{Evolution of temperature distributions for the disk model
with $\Sigma_0=10^3$~g~cm$^{-2}$. Left top panel: midplane temperature for
20~yr. Left bottom panel: disk surface temperature. Top right panel:
temperature at half surface density from the upper boundary of the disk
to the equator. Right bottom panel: long-term evolution of midplane
temperature over 1000~yr. The black curve shows the dependence of the
characteristic thermal time $t_{\rm therm}$ on distance.}
\label{fig_04}
\end{figure*}

Fig.~\ref{fig_03} shows the distributions of physical quantities for the
model with enhanced surface density, $\Sigma_0=10^3$~g~cm$^{-2}$, at the
moment of 10~yr from the beginning of evolution. The structure of this
disk is inhomogeneous. Two perturbations can be distinguished in the
energy density distribution of IR radiation: in the vicinity of 10 and
40~au. The morphology of these disturbances generally repeats the pattern
described for the previous model. However, unlike the previous model, the
heated area in the vicinity of disturbances does not reach the midplane.
This can be clearly seen from the top left panel of Fig.~\ref{fig_04},
which shows evolution of the midplane temperature distribution in the
first 20~yr. The waves do not appear in this diagram, with the exception
of weak oscillations in vicinity of 3~au. The inward propagating waves,
however, are clearly visible on surface temperature distributions (lower
left panel of Fig.~\ref{fig_04}) and temperatures at half surface density
up to equator (upper right panel of Fig.~\ref{fig_04}). Over the
considered time interval, the disturbances are generated in the vicinity
of 20~au and spread inward in within 3~yr. As they approach the inner
boundary of the disk, the propagation velocity of perturbations
decreases. The period of passage of these waves is much shorter than the
thermal time scale $t_{\rm therm}$ calculated for the entire thickness of
the disk. This is evident from the bottom right panel of
Fig.~\ref{fig_04}, where the distribution of thermal time scale for the
entire thickness of the disk is plotted. The time $t_{\rm therm}$ is
about hundreds of years for the region between 3 and 20~au. On the
same panel we show the evolution of midplane temperature of the disk
during 1000~yr, i.e. over time comparable with the characteristic thermal
time. In this diagram, the waves inside 20~au are also not
identified but quasi-periodic perturbations propagating inward from the
outer edge of the disk are visible with a occurrence period
$\approx$100~yr. Obviously, the waves in the outer region of the disk
are identified on midplane temperature distribution due to the fact that
the characteristic thermal times in this region are significantly shorter
than in more internal parts, i.e. disk being perturbed on the surface in
these regions has time to warm up to the equator.

\section{Discussion}
The main purpose of this work is to test the possibility of spontaneous
excitation of surface thermal waves in protoplanetary disks. To achieve
this goal, we updated the previously developed 1+1D disk model with a
more accurate (two-dimensional) calculation of the disk heating by the
stellar radiation. In the modified model, heating is calculated by
integrating the radiative transfer equation along the entire disk taking
into account the radial density gradient inside the cell, for which the
heating function is calculated. This allows us to account for geometric
effects that are of key importance for this task. The simulation results
showed that the instability does indeed arise, leading to the emergence
of inward moving perturbations. Thus, we confirm the previously obtained
by other authors conclusions on possible spontaneous excitation of the
surface thermal waves in protoplanetary disks.

In our disk model, we focuse on the occurrence of thermal surface waves,
as far as feasible ``in a pure form'' in connection with the
fact that there are still questions of their formation. Therefore, many
processes essential for disk evolution are not taken into account in this
model. In particular, we do not take into account internal viscous
heating, which can play an important role in the evolution of the disk,
providing an irregular mode of accretion (we studied this effect earlier
in \cite{2020ARep...64....1P, 2020ARep...64..815M}. Neglecting the
viscous heating is equivalent to assuming low accretion rate through the
disk.

Surely, at temperatures of the order of thousands of degrees, the
assumption of our model that the opacity is due only to
dust is not fulfilled. At such temperatures dust evaporation becomes
important, and gas begins to dominanate in opacity.
At high temperatures other processes become significant, such as the
dissociation of hydrogen. In this case, one can expect the appearance of
various instabilities in the disk and the complex dynamics associated
with them. This is a separate issue requiring detailed study. We note
that with our model parameters such high temperatures are
obtained only in the most inner ($<3$~au) regions of the disk and do not
affect more distant regions, where the thermal waves originate and
begin to propagate.

In our model, we also do not take into account dust sedimentation,
migration, growth and destruction, which play a critical role in
protoplanetary disks. Taking this processes into account can
significantly modify or even completely suppress the instability of the
disk associated with the effects of self-shadowing. We also note that the
mere presence of dust leads to its own instabilities, such as streaming
instability~\citep{2005ApJ...620..459Y}. Streaming instability now
becomes very popular for explaining turbulence and formation of planetary
embryos. The study of these processes is a separate direction for which
complex dynamical models are developed.  In this work, we do not consider
all these processes, postponing the study of their interaction with the
irradiation instability for the future.

A significant difference between our results and the conclusions of the
analytical model by \cite{2021ApJ...923..123W} are
characteristic timescales for propagation of perturbations. According to the
analytical model presented in \cite{2021ApJ...923..123W},
characteristic propagation time of perturbations corresponds to the
thermal time scale for the entire vertical column of the disk. This
result is natural, since in frame of that analytical model, the height
of the disk is determined exclusively by the midplane temperature. In our
model, the propagation times for optically thick disks turn out to be
significantly smaller than the thermal time scale $t_{\rm term}$. This is
explained by the fact that traveling perturbations do not affect the
entire thickness of the disk, i.e. wave excitation mechanism can work in
the sub-surface layers.

The obtained results indicate the need to study the irradiation
instability using more consistent models. Indeed, the 1+1D approach we
used is based on several key approximations that can significantly
distort the real picture. Such approximations are: 1) the hydrostatic
equilibrium in the vertical direction; 2) the absence of the thermal
radiation diffusion in the radial direction; 3) the lack of dynamical
interaction between disk regions in the radial direction. The influence
of these effects on the excitation of surface thermal waves should be
explored within the framework of a two-dimensional or three-dimensional
hydrodynamic models with a detailed calculation of the  radiative
transfer. Particular attention in such model should be paid to the
treatment of the heating function by the stellar radiation.

\section*{ACKNOWLEDGMENTS}
The authors thank the reviewer for valuable comments
and constructive suggestions for improving this article.

\section*{Funding}
The reported study was funded by RFBR according to the research project
20-32-90103. VVA was supported by the Foundation for the Advancement of
Theoretical Physics and Mathematics "BASIS" (20-1-2-20-1).

\bibliographystyle{mn2e}
\bibliography{twave}

\end{document}